\documentclass[12pt]{elsart}
\usepackage{amstex}
\usepackage{amssymb}
\usepackage{epsfig}

\begin{document}

\begin{flushright}
BARI-TH 227/96 
\end{flushright}

\begin{frontmatter}

\title{Finite Size Analysis of the\\ U(1) Background Field Effective
Action}

\author[INFN,Dep]{P. Cea\thanksref{emcea}},
\author[INFN]{L. Cosmai\thanksref{emcosmai}} and
\author[INFN,Dep]{A. D. Polosa\thanksref{empolosa}}
\address[INFN]{INFN - Sezione di Bari - 
Via Amendola, 173 - I 70126 Bari - Italy} 
\address[Dep]{Dipartimento di Fisica Univ. Bari -
Via Amendola, 173 - I 70126 Bari - Italy} 
\thanks[emcea]{E-mail: cea@@bari.infn.it}
\thanks[emcosmai]{E-mail: cosmai@@bari.infn.it}
\thanks[empolosa]{E-mail: polosa@@bari.infn.it}

\begin{abstract} 
We apply the finite size scaling analysis to the derivative of the
density of the effective action for the lattice U(1) pure gauge theory
in an external constant magnetic field. We found the presence of a
continuous phase transition. Moreover, our extimate of of the critical
parameters gives values consistent with those extracted from the
analysis of the specific heat.
\end{abstract}
\end{frontmatter}

\section{Introduction}

In a recent paper~\cite{lavoro1} (henceforth referred as I) we proposed a new method
that allows to evaluate the gauge invariant effective action for
external background fields in lattice gauge theories. In I we
considered the so-called lattice Schr\"odinger
functional~\cite{Wolff86}:
\begin{equation}
\label{Eq1}
{\mathcal{Z}} \left[ \vec{A}^{\text{ext}} \right] = 
 \langle  \vec{A}^{\text{ext}} |  \exp(-HT) {\mathcal{P}} | 
\vec{A}^{\text{ext}}
\rangle  \, = \int {\mathcal{D}}U \exp(-S_W)
\end{equation}
where $S_W$ is the standard Wilson action.
In Equation~\eqref{Eq1} we integrate over the links $U_\mu(x)$ with the
constraints
\begin{equation}
\label{Eq2}
U_\mu(x)|_{x_4=0} = U_\mu^{\text{ext}} \,,
\end{equation}
where 
\begin{equation}
\label{Eq3}
U_\mu^{\text{ext}}(x) = {\text{P}} \exp \left\{ + iag  \int_0^1 dt \,
A_\mu^{\text{ext}}(x+ at {\hat{\mu}}) \right\} \,.
\end{equation}
The lattice effective action for the background field 
$\vec{A}^{\text{ext}}(\vec{x})$ is given by:
\begin{equation}
\label{Eq4}
\Gamma\left[ \vec{A}^{\text{ext}} \right] = -\frac{1}{T} 
\ln \left\{ \frac{{\mathcal{Z}}[U^{\text{ext}}]}{{\mathcal{Z}}(0)} \right\}
\end{equation}
where $T$ is the extension in the Euclidean time. In the continuum
where $T\rightarrow\infty$ it is easy to show that 
$\Gamma\left[ \vec{A}^{\text{ext}} \right]$ reduces to the vacuum
energy in presence of the background field
$\vec{A}^{\text{ext}}(\vec{x})$. Note that our effective action is by
definition gauge invariant. Moreover the effective action~\eqref{Eq4}
can be used for a non-perturbative investigation of the properties of
the quantum vacuum.

\section{U(1) in a constant background field}

In I we considered the four-dimensional U(1) lattice gauge theory in
an external constant magnetic field:
\begin{equation}
\label{Eq5}
A_k^{\text{ext}}(\vec{x}) = \delta_{k,2} x_1 B \,.
\end{equation}
The lattice links corresponding to the vector potential~\eqref{Eq5} are
\begin{equation}
\label{Eq6}
\begin{split}
U_2^{\text{ext}}(x) = \exp \left[ i a g B x_1 \right] 
= \cos(agBx_1) + i \sin(agBx_1) \,,\\
U^{\text{ext}}_1(x) =  U^{\text{ext}}_3(x) = U^{\text{ext}}_4(x) =  1 \,,
\end{split}
\end{equation}
where, due to the periodic boundary conditions, we have
\begin{equation}
\label{Eq7}
a^2 g B = \frac{2 \pi}{L_1} n^{\text{ext}}
\end{equation}
with $n^{\text{ext}}$ an integer, and $L_1$ the lattice extension in the
$x_1$ direction (in lattice units).

Due to the gauge invariance the effective action only depends on the constant
field strength $F^{\text{ext}}_{12}$. Thus we consider the density of the
effective action
\begin{equation}
\label{Eq8}
\varepsilon\left[ \vec{A}^{\text{ext}} \right] =
\frac{\Gamma\left[ \vec{A}^{\text{ext}} \right]}{V}=
-\frac{1}{\Omega} \ln \left[ 
\frac{{\mathcal{Z}}[U^{\text{ext}}]}{{\mathcal{Z}}(0)} \right] \,,
\end{equation}
where $\Omega=V \cdot T$, with $V$ the lattice spatial volume. In
order to avoid the problem of computing a partition function, we
focused on the derivative of  $\varepsilon\left[ \vec{A}^{\text{ext}}
\right]$  with respect to $\beta$:
\begin{equation}
\label{Eq9}
\varepsilon^{\prime} \left[ \vec{A}^{\text{ext}} \right] = 
\left\langle \frac{1}{\Omega} \sum_{x,\mu>\nu} \cos \theta_{\mu\nu}(x)
\right\rangle_0 -
\left\langle \frac{1}{\Omega} \sum_{x,\mu>\nu} \cos \theta_{\mu\nu}(x)
\right\rangle_{A^{\text{ext}}} \,.
\end{equation}
From our previous discussion it is clear that
$\varepsilon\left[ \vec{A}^{\text{ext}} \right]$ is the
vacuum energy density. Thus we expect that 
$\varepsilon^{\prime} \left[ \vec{A}^{\text{ext}} \right]$ behaves
like a specific heat. Indeed, in I we found that, near the critical
region $\beta \approx 1$,  
$\varepsilon^{\prime} \left[ \vec{A}^{\text{ext}} \right]$ displays a
maximum which increases by increasing the lattice volume. In addition,
the peak shrinks and shift towards $\beta=1$ by increasing
$L=\Omega^{1/4}$.

\section{Finite Size Scaling Analysis}

The aim of this paper is to determine the critical parameters by
applying the standard finite size scaling analysis~\cite{Barber89}.
Consider a finite system with effective linear dimension $L$. In
general the thermodynamical quantities depend on $\xi/a$ and $L/a$,
where $\xi$ is the correlation length and $a$ is the lattice spacing.
The finite size scaling hypothesis assumes that near the critical
point the microscopic length $a$ drops out.

Let us consider, for instance, the generalized susceptibility $\chi$.
In a  infinite system near the critical point we have
\begin{equation}
\label{Eq10} 
\chi \sim \xi^{\gamma/\nu} \,.
\end{equation}
In the finite geometry with size $L$, according to the finite size
scaling hypothesis, we can write
\begin{equation}
\label{Eq11}
\chi = \xi^{\gamma/\nu} \phi(\xi/L)
\end{equation}
where the scaling function $\phi(u)$ is smooth. Now, from 
$\xi \sim |T - T_c|^{-\nu}$ it follows
\begin{equation}
\label{Eq12}
\chi = L^{\gamma/\nu} \tilde{\phi} \left( L^{1/\nu} (T-T_c) \right)
\,,
\end{equation}
where $\tilde{\phi}(u)$ is another smooth scaling function. On a
finite system $\chi$ is supposed to be analytic in $T$. Thus
$\tilde{\phi}(u)$ must be a smooth function with a peak at some value
$u=u_0$. Near the peak we parametrize $\tilde{\phi}(u)$ as:
\begin{equation}
\label{Eq13}
\tilde{\phi}(u) = \frac{a_1}{a_2(u-u_0)^2+1}
\end{equation}
with $a_1$, $a_2$, $u_0$ positive constants.

From Equations~(\ref{Eq12}) and~(\ref{Eq13}) we obtain easily:
\begin{equation}
\label{Eq14}
\chi = \frac{a_1 L^{\gamma/\nu}}{a_2 L^{2/\nu} \left[ T - T^{*}(L)
\right]^2 + 1} 
\end{equation}
where
\begin{equation}
\label{Eq15}
T^{*}(L) = T_c + u_0 L^{-1/\nu} \,.
\end{equation}
Then, if we plot the susceptibility versus the temperature, we see a
peak at $T^{*}(L)$ with a width $L^{-1/\nu}$ and height 
$\sim L^{\gamma/\nu}$. Further, the critical exponents $\gamma$ and
$nu$ are related by the so-called hyperscaling~\cite{Barber89} 
relation~\cite{Barber89} 
\begin{equation}
\label{Eq16}
\gamma = 2 - \nu d
\end{equation}
where $d$ is the dimensionality of the system. Equation~\eqref{Eq16}
implies that
\begin{equation}
\label{Eq17}
\frac{1}{\nu} = \frac{d}{2}+ \frac{1}{2}\frac{\gamma}{\nu}   \,.
\end{equation}
These considerations can be directly applied to our generalized
susceptibility  $\varepsilon^{\prime} \left[ \vec{A}^{\text{ext}}
\right]$. In order to determine the critical parameters we need to
simulate our gauge system on lattices with different sizes. However we
must take fixed the external magnetic field, for
$\varepsilon^{\prime} \left[ \vec{A}^{\text{ext}} \right]$ depends
even on $gB$. Thus, according to Eq.~\eqref{Eq7}, we must work with
$L_1$ and $n^{\text{ext}}$ fixed.
As a consequence we perform the numerical
simulations on lattices with sizes $L_1=64$, $L_2=L_3=L_4\equiv L=6,
8, 10, 12, 14$ and $n^{\text{ext}}=2$. It is worthwhile to stress that
the contributions to  $\varepsilon^{\prime} \left[
\vec{A}^{\text{ext}} \right]$  due to the frozen time slice at
$x_4=0$ and to the fixed boundary conditions at the lattice spatial
boundaries must be subtracted. Indeed, the physically relevant
quantity is due to the ``internal'' links, i.e. only the dynamical
links must be taken into account in evaluating 
$\varepsilon^{\prime} \left[\vec{A}^{\text{ext}} \right]$.
In Figure~1 we display the derivative of the internal energy
$\varepsilon^{\prime} \left[\vec{A}^{\text{ext}} \right]$ versus
$\beta$ for the $64 \times 12^3$ lattice. Note that in the strong
coupling region $\beta \lesssim 0.7$  
$\varepsilon^{\prime} \left[\vec{A}^{\text{ext}} \right]$ vanishes so
that the density of the effective action can be recovered by
integrating over $\beta$ without any subtraction.

In Figure~2 we show
$\varepsilon^{\prime} \left[\vec{A}^{\text{ext}} \right]$ near the
critical region for the different lattice sizes. We try a fit of the
peak according to Eq.~\eqref{Eq13}:
\begin{equation}
\label{Eq18}
\varepsilon^{\prime} \left[\vec{A}^{\text{ext}} \right] =
\frac{a_1(L_{\text{eff}})}{a_2(L_{\text{eff}}) 
[ \beta - \beta^{*}(L_{\text{eff}})]^2 +1}
\end{equation}
where the effective size of the lattice is~\cite{Barber89}
\begin{equation}
\label{Eq19}
L_{\text{eff}} = \Omega_{\text{int}}^{1/4}  \,.
\end{equation}
In Equation~\eqref{Eq19} $\Omega_{\text{int}}$ is the internal lattice
volume, i.e. the volume occupied by the dynamical lattice sites. 
If we denote by $\Omega_{\text{ext}}$ the lattice sites whose links
are fixed according to Eq.~\eqref{Eq6}, thus we get:
\begin{equation}
\label{Eq20}
\Omega_{\text{ext}}= L_1 L_2 L_3 + 
(L_4-1)(L_1 L_2 L_3 - (L_1-2)(L_2-2)(L_3-2)) \,,
\end{equation}
where the first term is the volume of the frozen time slice $x_4=0$
while the second term is due to the fixed links at the lattice spatial
boundaries.
Whereupon we have:
\begin{equation}
\label{Eq21}
\Omega_{\text{int}} = L_1 L_2 L_3 L_4 -
\Omega_{\text{ext}} = \Omega -\Omega_{\text{ext}}  \,.
\end{equation}
It turns out that our lattice data are in satisfying agreement with
Eq.~\eqref{Eq18}. Indeed near the pseudocritical coupling we fit the
data according to Eq.~\eqref{Eq18} by means of the standard MINUIT
subroutine of the CERN library. We find rather good fits (solid lines
in Fig.~2) with $\chi^2/{\text{d.o.f.}} \sim 1$.
As a result of the fits we get the values
of the three parameters for the five different values of $L_{\text{eff}}$.
According to the finite size scaling hypothesis $a_1(L_{\text{eff}})$
should behave as
\begin{equation}
\label{Eq22}
a_1(L_{\text{eff}}) = c  L_{\text{eff}}^{\gamma/\nu}  \,.
\end{equation}
In Figure~3 we report $a_1(L_{\text{eff}})$ versus $L_{\text{eff}}$
together with the result of the fit Eq.~\eqref{Eq22}.
We find
\begin{align}
\label{Eq23}
c & =   0.63 (24)  \\ \nonumber
\frac{\gamma}{\nu} & =  0.84 (14) 
\end{align}
with $\chi^2/{\text{d.o.f.}} = 1.2$. The pseudocritical couplings are fitted
according to Eq.~\eqref{Eq15}:
\begin{equation}
\label{Eq24}
\beta^*(L_{\text{eff}}) = \beta_c + d L_{\text{eff}}^{-1/\nu} \,.
\end{equation}
We find (see Fig.~4):
\begin{align}
\label{Eq25}
\frac{1}{\nu} & = 3.48 (65) \\ \nonumber
\beta_c & = 1.0089 (47)
\end{align}
with $\chi^2/{\text{d.o.f.}} = 1.7$. The values Eqs.~\eqref{Eq23}
and~\eqref{Eq25} for the critical exponent are quite consistent with the
hyperscaling relation Eq.~\eqref{Eq16}. Indeed from Eqs.~\eqref{Eq17}
and~\eqref{Eq23} we find
\begin{equation}
\label{Eq26}
\frac{1}{\nu} = 2.42 (7) \,,
\end{equation}
which is compatible with Eq.~\eqref{Eq25} within two standard deviations.

Finally, in Fig.~5 we check that the universality law in the critical
region:
\begin{equation}
\label{Eq27}
L_{\text{eff}}^{-\gamma/\nu} \varepsilon_{\text{int}}^{\prime} = \left[
\vec{A}^{\text{ext}} \right] = \tilde{\phi} \left[
L_{\text{eff}}^{1/\nu}(\beta-\beta_c) \right] 
\end{equation}
holds quite well for the lattices with $L=10$, $12$, $14$
corresponding to $L_{\text{eff}}=13.75$, $16.16$, $18.46$
respectively.

\section{Conclusions}

We would like to emphasize that our extimation of the infinite volume critical
coupling is in perfect agreement with the values extracted from the specific
heat~\cite{Lautrup80,Lippert93}. However it is well established that on
lattices with periodic boundary conditions, the U(1) phase transition is
exhibiting a weakly first-order nature~\cite{Lippert93}. On the other hand, our
data are consistent with a continuous phase transition. As a matter of fact, it
turns out that the phase transition of compact U(1) pure lattice gauge theory
with the Wilson action on lattices with closed topology~\cite{Jersak95,Lang94}
or with fixed boundary conditions~\cite{Baig94} is continuous. Moreover our
extimation of the critical exponents is consistent with the results of
Ref.~\cite{Jersak95}. Thus, the finite size scaling analysis discussed in this
paper results in a successfully quantitative test of our proposal of a
gauge invariant effective action on the lattice.

\clearpage

\section*{FIGURE CAPTIONS}

\renewcommand{\labelenumi}{Figure \arabic{enumi}.}
\begin{enumerate}
\item  
The derivative of the energy density due to links belonging to
$\Omega_{\text{int}}$ versus $\beta$ for the $64 \times 12^3$ lattice.
\item 
$\varepsilon_{\text{int}}^{\prime}$ near the critical region for the
lattices 
$64 \times 6^3$ (open circles),  $64 \times 8^3$ (open squares), 
$64 \times 10^3$ (open triangles),   $64 \times 12^3$ (open diamonds), and
$64 \times 14^3$ (open crosses), with the superimposed fits
Eq.~\eqref{Eq18}.
\item
The maximum of $\varepsilon_{\text{int}}^{\prime}$ obtained from the fit
Eq.~\eqref{Eq18} versus $L_{\text{eff}}$ with superimposed the fit
Eq.~\eqref{Eq22}.
\item
The pseudocritical couplings at various lattice sizes with
superimposed the fit Eq.~\eqref{Eq24}.
\item
The universality law Eq.~\eqref{Eq27} in the critical region. Symbols
as in Fig.~2.
\end{enumerate}

\newpage
\begin{figure}[t]
\begin{center}
\epsfig{file=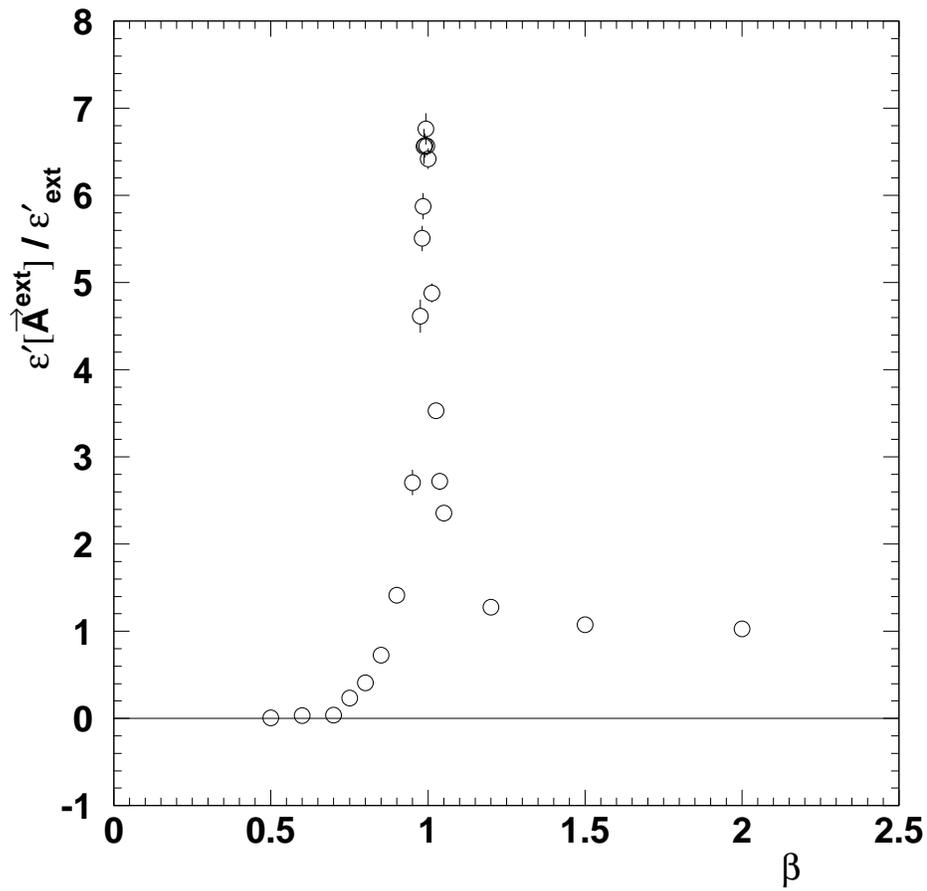,width=\textwidth}
\end{center}
\vspace{-10pt}
\caption{
The derivative of the energy density due to links belonging to
$\Omega_{\text{int}}$ versus $\beta$ for the $64 \times 12^3$ lattice.
}
\label{Fig:1}
\end{figure}
\newpage
\begin{figure}[t]
\begin{center}
\epsfig{file=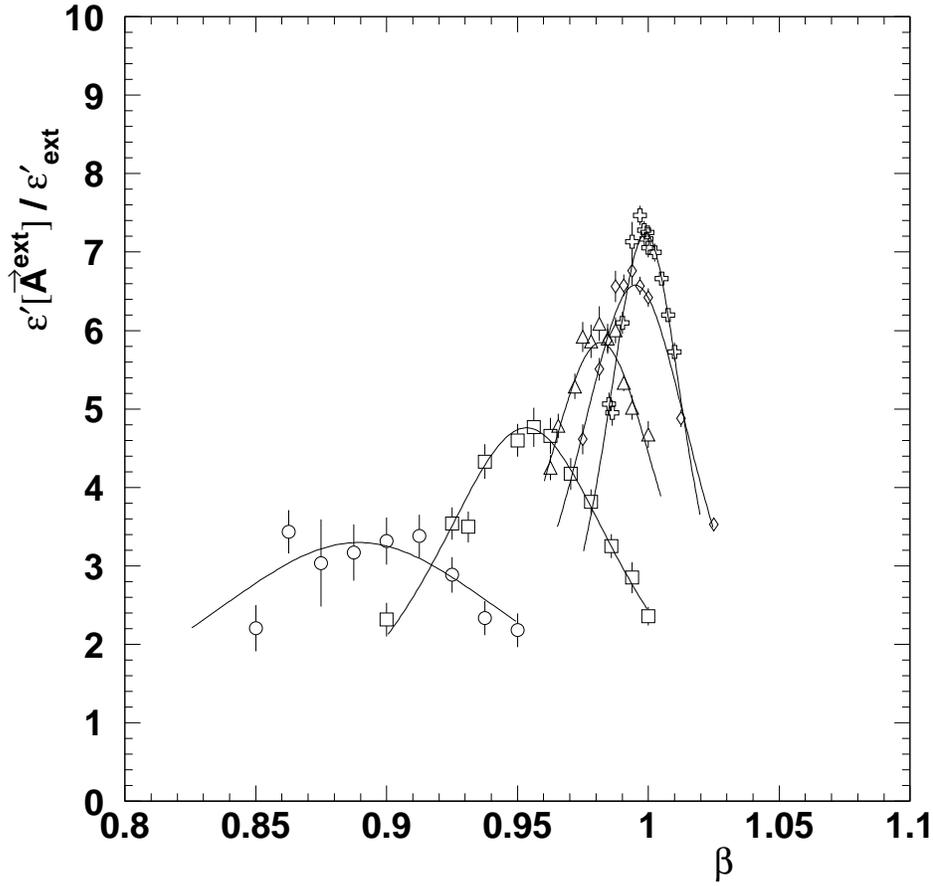,width=\textwidth}
\end{center}
\vspace{-10pt}
\caption{
$\varepsilon_{\text{int}}^{\prime}$ near the critical region for the 
lattices
$64 \times 6^3$ (open circles),  $64 \times 8^3$ (open squares), 
$64 \times 10^3$ (open triangles),   $64 \times 12^3$ (open diamonds), and
$64 \times 14^3$ (open crosses), with the superimposed fits
Eq.~\eqref{Eq18}.
}
\label{Fig:2}
\end{figure}
\newpage
\begin{figure}[t]
\begin{center}
\epsfig{file=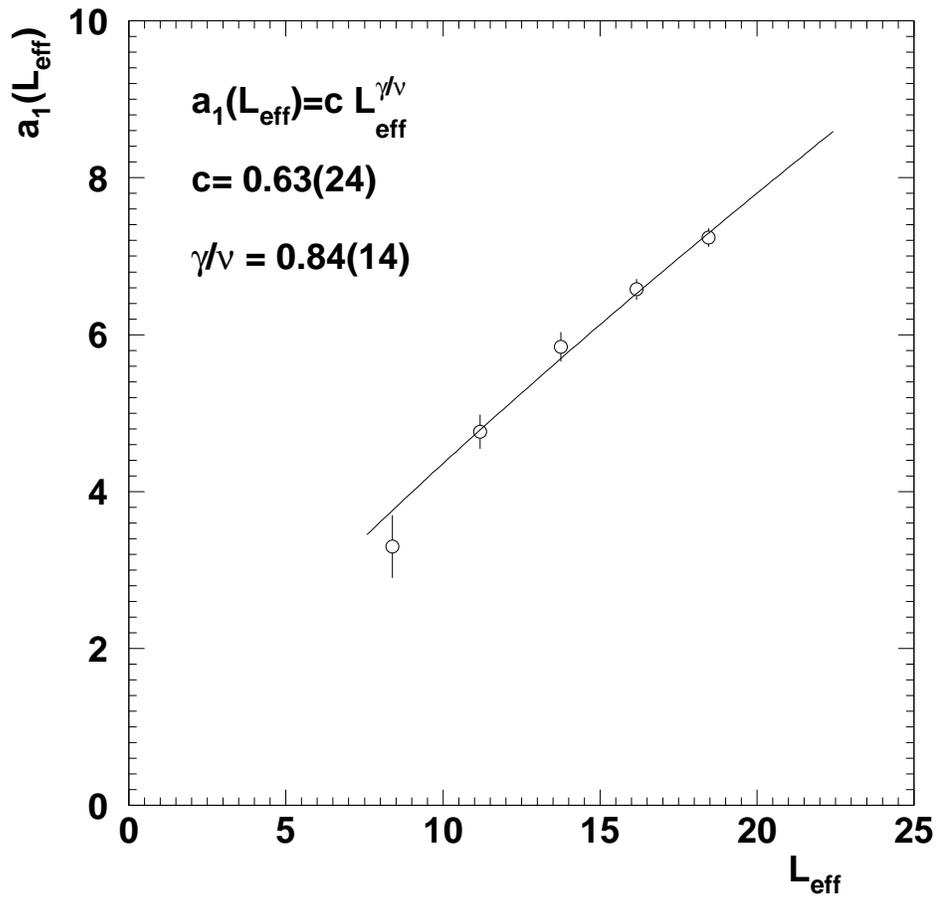,width=\textwidth}
\end{center}
\vspace{-10pt}
\caption{
The maximum of $\varepsilon_{\text{int}}^{\prime}$ obtained from the fit
Eq.~\eqref{Eq18} versus $L_{\text{eff}}$ with superimposed the fit
Eq.~\eqref{Eq22}.
}
\label{Fig:3}
\end{figure}
\newpage
\begin{figure}[t]
\begin{center}
\epsfig{file=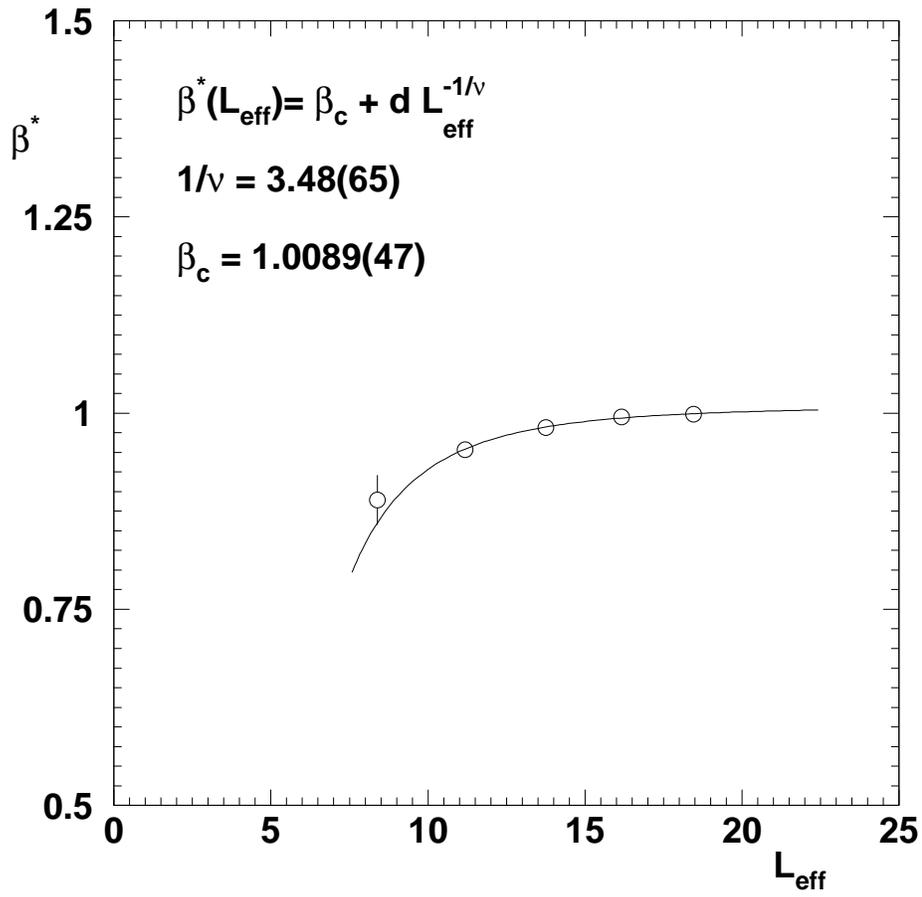,width=\textwidth}
\end{center}
\vspace{-10pt}
\caption{
The pseudocritical couplings at various lattice sizes with
superimposed the fit Eq.~\eqref{Eq24}.
}
\label{Fig:4}
\end{figure}
\newpage
\begin{figure}[t]
\begin{center}
\epsfig{file=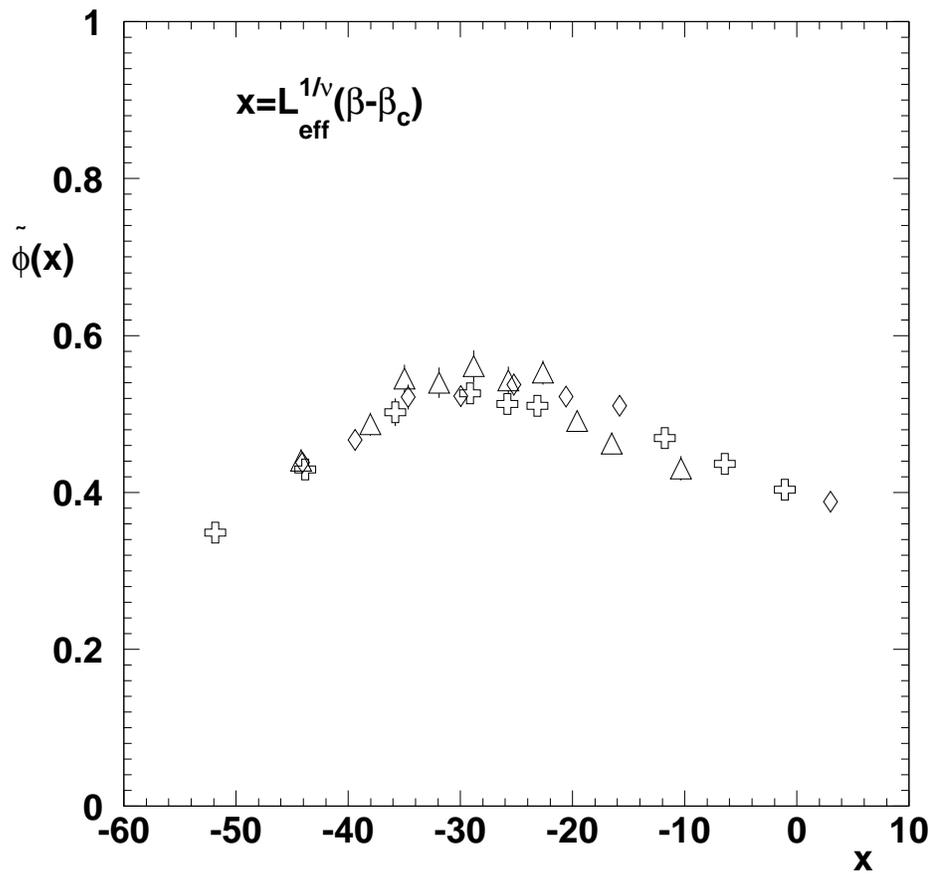,width=\textwidth}
\end{center}
\vspace{-10pt}
\caption{
The universality law Eq.~\eqref{Eq27} in the critical region. Symbols
as in Fig.~2.
}
\label{Fig:5}
\end{figure}

%%%%%%%%%%%%%%%%%%%%%%%%%%%%%%%%%%%%%%%%%%%%%%%%%%%%%%%%%%%%%%%%%%%%%%%%%%%%%%%%
\end{document}